# Artificial Intelligence Enabled Reagent-free Imaging Hematology Analyzer

Xin Shu,[1,#] Sameera Sansare,[2,3,#] Di Jin,[4] Xiangxiang Zeng,[5] Kai-Yu Tong,[1] Rishikesh Pandey,[3,6,*] Renjie Zhou,[1,7,*]

[1]Department of Biomedical Engineering, The Chinese University of Hong Kong, Shatin, New Territories, Hong Kong SAR, China
[2]Department of Pharmaceutical Sciences, University of Connecticut, Storrs, Connecticut, 06269, USA
[3]Connecticut Children's Innovation Center, University of Connecticut School of Medicine, Farmington, Connecticut, 06032, USA
[4]Computer Science and Artificial Intelligence Laboratory, Massachusetts Institute of Technology, Cambridge, Massachusetts 02139, USA
[5]School of Information Science and Engineering, Hunan University, Changsha 410076, China
[6]Department of Biomedical Engineering, University of Connecticut, Storrs, Connecticut, 06269, USA
[7]Shun Hing Institute of Advanced Engineering, The Chinese University of Hong Kong, Shatin, New Territories, Hong Kong SAR, China
#: Authors contributed equally; Corresponding Authors' e-mail addresses: rishikesh.pandey@uconn.edu, rjzhou@cuhk.edu.hk

**Abstract**: Leukocyte differential test is a widely performed clinical procedure for screening infectious diseases. Existing hematology analyzers require labor-intensive work and a panel of expensive reagents. Here we report an artificial-intelligence enabled reagent-free imaging hematology analyzer (AIRFIHA) modality that can accurately classify subpopulations of leukocytes with minimal sample preparation. AIRFIHA is realized through training a two-step residual neural network using label-free images of separated leukocytes acquired from a custom-built quantitative phase microscope. We validated the performance of AIRFIHA in randomly selected test set and cross-validated it across all blood donors. AIRFIHA outperforms current methods in classification accuracy, especially in B and T lymphocytes, while preserving the natural state of cells. It also shows a promising potential in differentiating CD4 and CD8 cells. Owing to its easy operation, low cost, and strong discerning capability of complex leukocyte subpopulations, we envision AIRFIHA is clinically translatable and can also be deployed in resource-limited settings, e.g., during pandemic situations for the rapid screening of infectious diseases.

## Introduction

Leukocytes play an important role in maintaining the normal function of human immune systems. For instance, B and T lymphocytes can produce antibodies to defend the body against foreign substances, such as bacteria and viruses. Abnormal leukocyte differential counts are indications of malfunctions of the immune system or infectious diseases[1]. For instance, a sharp increase in neutrophil-to-lymphocyte ratio serves as an independent risk factor for SARS-CoV-2 infection[2-4]. To differentiate basic leukocyte types, volume and granularity parameters are often measured through electrical impedance and light scattering-based cytometry techniques[5]. For more complex leukocyte types with similar morphologies (e.g., B and T lymphocytes), fluorescent molecules bound with antibodies that specifically target the proteins expressed on the surface are typically used to activate fluorescence emission which can be captured by detectors for population counting. Although antibody labeling based flow cytometry methods are widely used in the clinical laboratories, there remains a few drawbacks. Firstly, the chemical labeling process may affect the original cell physical state and viability that could affect the detection accuracy[6]. Secondly, an extensive list of expensive reagents is required for differentiating more cell types. Thirdly, the measured labelled cells cannot be reused for any further testing. Finally, dyes are susceptible to photobleaching which can affect long-term observation results.

Label-free imaging methods can potentially solve the aforementioned issues[3,7-10]. For instance, a hemogram based on Raman imaging has been proposed to discern leukocytes[11]. While this innovative approach leverages the unique biochemical attributes for the classification, it is limited by the weak spontaneous Raman signal, thus not suitable for high-throughput applications in a clinical setting. Quantitative phase microscopy (QPM) is a rapidly emerging imaging modality that is suitable for high-speed imaging of unlabeled specimens. In QPM, the exact optical path-length delay associated with the density and thickness at each point in the specimen is mapped, which has enabled label-free imaging of transparent structures (e.g., live cells) with a high imaging contrast [12-14]. In recent years, QPM has been used for single-cell analysis by extracting quantitative biomarkers, e.g., measuring cell dry mass to quantify cell growth[15,16], studying red blood cell rheology[17,18], characterizing cell viability[19], analyzing large cell population[20,21], and screening cancer[22], etc. However, most studies have primarily relied on interpreting the QPM results in terms of a few principal morphological characteristics. Recently, several laboratories including our own have sought to shift the paradigm by utilizing machine learning (ML) and artificial intelligence (AI) for analyzing and interpreting QPM data[23-25]. The full field and fast imaging attributes of QPM enable availability of volumes of high-dimension imaging and therefore make QPM a unique modality for the application of ML/AI to those tasks involving cell classification and imaging.

With recent developments in ML/AI, e.g., visual geometry group (VGG)[26], inception[27], and residual neural network (ResNet)[28,29], abundant training data is available to train a model to extract important image features to classify targeted objects [30,31]. Compared with previous manual feature extraction analysis methods, the new approaches in ML/AI may offer features with statistically significant higher sensitivity and specificity. Among the recent ML/AI methods, ResNet tackles the gradient vanishing problem by creating shortcut paths to jump over layers. Conversion among different types of biomedical images and the segmentation of certain cell structures have been achieved by using ResNet building blocks[32-34]. With such exciting developments, ML/AI have been applied to label-free imaging cytometry systems to tackle complicated cell analysis problems[21,24,35]. For instance, machine learning for the differentiation of B and T lymphocytes has been achieved on bright-field and dark-field microscopy platforms[36]. To further improve the detection accuracy and specificity of leukocyte subtypes, 3D QPM techniques has been proposed and demonstrated [37,38].

In this work, we propose a rapid, low-cost AI-enabled reagent-free imaging hematology analyzer (AIRFIHA) that can classify complex leukocyte types in human blood samples. AIRFIHA is based on leveraging the morphological attributes of phase images from a custom-built QPM system and a cascaded-ResNet for leukocyte classification. From this proof-of-principle study on six human donors, we have achieved a classification accuracy of 90.5% on average for monocytes, granulocytes, and B and T lymphocytes. The robustness and applicability of our proposed method have been confirmed by conducting cross-donor validation experiments. We further investigated the potential of AIRFIHA in discerning human CD4 and CD8 T cells. AIRFIHA demonstrated a much higher accuracy when compared with methods based on negative isolated leukocyte classification and a comparable or slightly better accuracy when compared with methods based on positive isolated leukocyte classification. This study shows a promising perspective when applying AIRFIHA for automated clinical blood testing applications, which is especially useful in resource-limited settings and during pandemic situations.

# Results

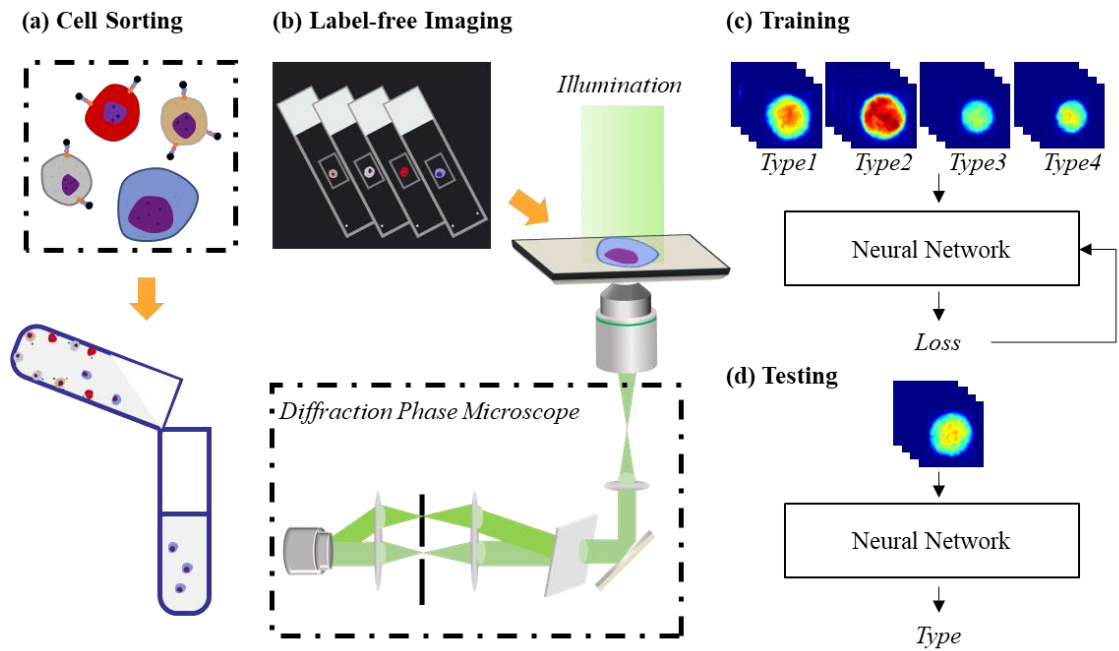

**Fig. 1 Workflow of the AIRFIHA system**. **a**, Different types of leukocytes are negatively separated using antibody-labeled magnetic particles. **b**, A diffraction phase microscope is used for obtaining quantitative phase images of sorted leukocytes. **c**, Deep learning model is trained for classifying the leukocyte types. **d** A trained neural-network model is used to predict leukocyte types.

## AIRFIHA system

In this work, the classification of human leukocyte types is achieved using a QPM system and a neural network, as conceptually illustrated in Fig. 1. The exact configuration of the QPM system is based on a diffraction phase microscope (DPM)[2,39,40], which can provide highly stable and accurate phase imaging of cells. The imaging resolution of the QPM system is 590 nm, while the field of view is around 61 μm x 49 μm. Compared with optical diffraction tomography[37,38], QPM does not necessitate a complex imaging system and expensive computation requiring a large amount of data, and the system is relatively cost-effective with a smaller footprint. The leukocyte samples were isolated from the fresh blood samples of six healthy donors, and the blood sample used for the leukocyte separation for each donor was in 1-3 ml range, depending on the minimum volume requirement as per manufacturer's instruction for the leukocyte subpopulations.. The leukocytes were negatively isolated by using antibody labeled magnetic particles as illustrated in Fig. 1 (a) (refer to detailed sorting procedure in "Methods"). Then, the isolated sample was diluted in PBS (phosphate buffer saline) and mounted between two glass coverslips before placing it onto a home-built QPM system as illustrated in Fig. 1(b) (refer to the detailed sample preparation procedure in "Methods"). Phase images of each leukocyte type were retrieved from the measured interferograms (refer to the detailed description of the QPM system and the phase retrieval method in **Supplementary**). After thousands of phase images of labeled leukocytes of different types were measured, all the leukocytes in each phase image were segmented to construct the training and testing dataset[41]. A neural network was constructed, trained, and validated for classifying the leukocytes using the phase image dataset (Fig. 1c). A detailed description of the neural network is provided in the following section. Finally, the AIRFIHA system was used to identify leukocyte types of new samples (Fig. 1d).

## Leukocyte classification method

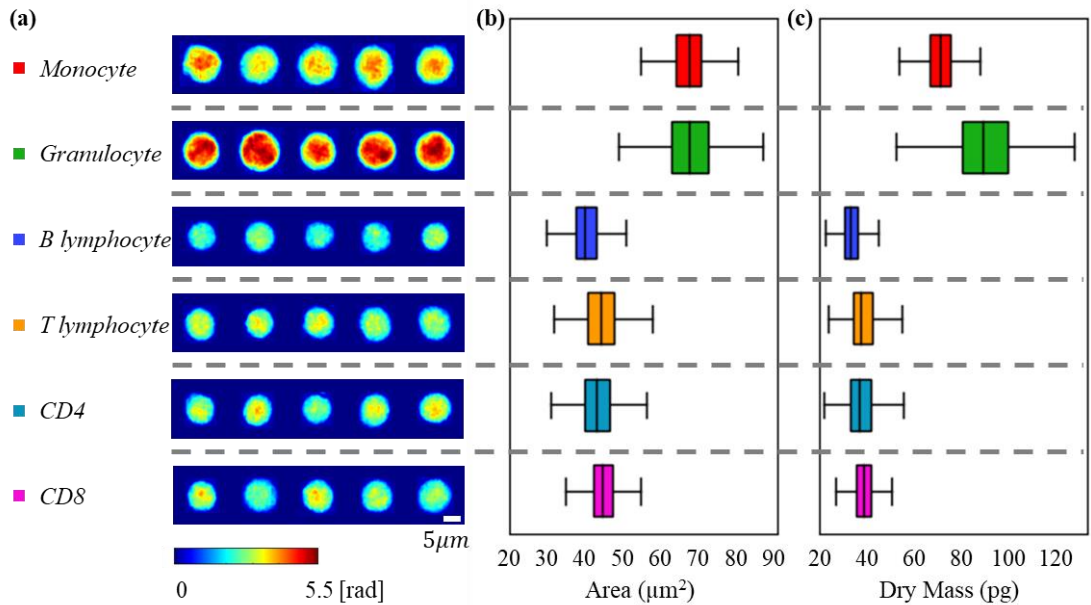

**Fig. 2 Representative phase images, area, and dry mass distributions of different leukocyte types**. **a**, Representative phase images for each leukocyte type. **b**, **c** Area and dry mass distributions for all the leukocyte types in the dataset. Monocytes (red), granulocytes (green), B lymphocytes (blue), T lymphocytes (orange), CD4 cells (light blue) and CD8 cells (purple). Scalebar: 5 μm.

Phase maps of labeled leukocytes of four different types from multiple donors were measured to construct the main dataset, including 857 monocytes, 738 granulocytes, 700 B lymphocytes, and 821 T lymphocytes (i.e., 1521 lymphocytes in total). Additionally, we had a phase map dataset for two subtypes of T lymphocytes, containing 211 CD4 cells and 220 CD8 cells. Representative phase maps for each leukocyte subtype are shown in Figure 2a. Based on these phase maps, area and dry mass distributions were generated for all the leukocyte types (Fig. 2b, c). Note that cell dry mass quantifying the total protein content in a cell can be precisely determined from the phase map, and it has been well explored for cell phenotyping[15,42]. As shown in Fig. 2b, c, monocytes and granulocytes have similar areas but very different dry masses (p-value < 0.001), while they are well separated from all the other lymphocytes (B and T lymphocytes and CD4 and CD8 cells) through both area and dry mass distributions (p-value < 0.001). For the main subtypes of lymphocytes, i.e., B and T lymphocytes, they are different in both cell area and dry mass (p-value <0.001), but the differences are small. The subtypes of T lymphocytes, i.e., CD4 and CD8 cells, have similar cell dry mass and slightly different cell area distributions (p-value <0.001).

To achieve better detection specificity and accuracy for classifying leukocyte types of similar morphology, we will fully explore the quantitative phase image information that contains more cell features. We first constructed a neural network by cascading two ResNets as shown in Fig. 3a, b. This neural network was designed to simultaneously classify monocytes, granulocytes, and B and T lymphocytes using a two-step classification routine. The leukocyte types in these two classifiers are allotted in a way that each leukocyte type within one classifier share similar degrees of classification difficulties. The first ResNet (Fig. 3a) is used to classify monocytes, granulocytes, and lymphocytes. The predicted lymphocytes are then put into the second ResNet (Fig. 3b) for further classification into B and T lymphocytes. Due to the similarity of these two classification tasks, the second ResNet was fine-tuned from the first ResNet. ResNets of different depths were explored, while the highest validation accuracy was obtained on the ResNet-10 that had around 1.5 million trainable parameters. ResNet-10 has 10 layers, i.e., one input convolution layer, 8 convolution layers from 4 building blocks (each building block has 2 convolution layers), and one final dense layer. The shortcut connects the head and tail of each building block, which helps to restore the crucial shallower features for prediction. The layer size is halved, and the kernel quantity is doubled for every 1, 2, 1 building blocks.

Batch normalization (Batch Norm)[43] is applied for each mini-batch after each convolutional layer. Rectified Linear Unit (Relu)[44] is used as the nonlinear activation function. After the last building block, an average pool and a flatten layer are applied to convert each two-dimensional feature map into one value, thus for 256 feature maps, a $256 \times 1$ vector is obtained to represent each of the input images. Probabilities of each type are produced based on this feature vector via a dense layer with the Softmax activation function[45]. For the monocyte-granulocyte-lymphocyte classification task, probabilities of these three types are produced, while for B-T lymphocyte classification, two probability values are produced. The type with the largest probability value is used to make the final decision. To explore the differentiation capability of CD4 and CD8 cells, a separate ResNet was trained by fine-tuning the B-T lymphocyte classifier for the new classification task (Fig. 3c). Details on the training and validation of the classification model are provided in the **Method** section.

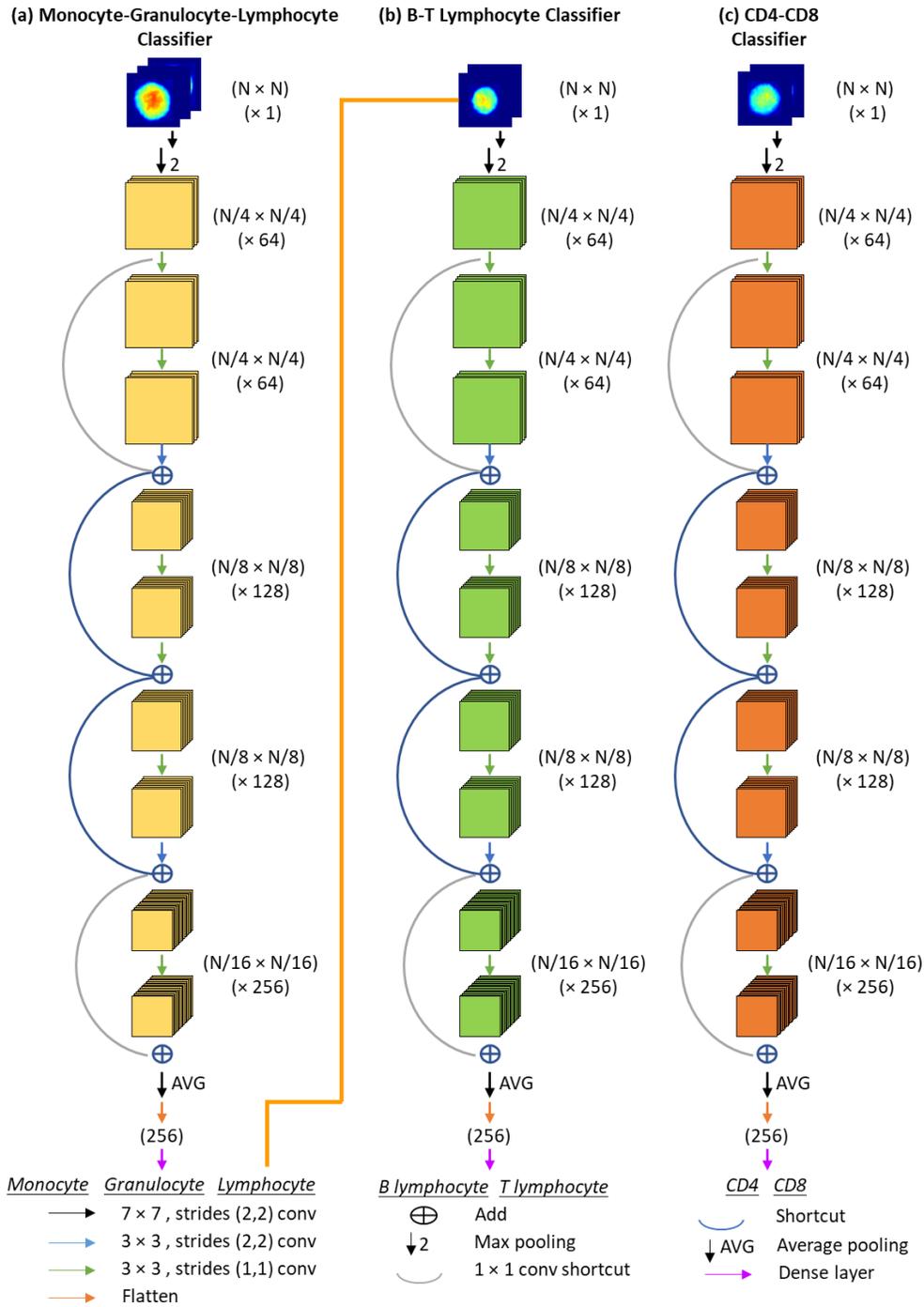

**Fig. 3 The ResNets constructed for classification of leukocytes**. **a, b** The cascaded-ResNet-10 for classifying monocytes, granulocytes, and B and T lymphocytes. **c**, The ResNet-10 for classifying CD4 and CD8 cells.

**Classification results**

To test the classification capability of our AIRFIHA system, a test set was first constructed by randomly selecting 100 cells from four leukocytes, i.e., monocytes, granulocytes, and B and T lymphocytes. Notably, the test set was not contained in the training set. The classification results were evaluated using recall, precision, and F1 score[46]. F1 score, which is the harmonic mean of recall and precision, is used to characterize the final classification result. The F1 scores from the first classifier for monocytes, granulocytes, and lymphocytes are 94%, 95.4%, and 97.7%, respectively (detailed numerical values for recall, precision, and F1 are provided in Table S1). The F1 scores from the second classifier for B and T lymphocytes are 88.2% and 88.8%, respectively (detailed numerical values for recall, precision, and F1 are provided in Table S2). The overall detection results are summarized and visualized in Fig. 4a and Table S3. The precision-recall curves[47] for each of the classifier in the cascaded-ResNet are plotted and shown in Fig. 4b and 4c. The values of the area under the precision-recall curve (AUPRC) for lymphocytes, monocytes, and granulocytes in the first classifier are 1.00, 0.98 and 0.98, respectively. The values of AUPRC for B and T lymphocytes in the second classifier are 0.96 and 0.94, respectively. Our B/T cell classification accuracy is comparable with the method based on 3D quantitative phase imaging [38] (note that leukocytes here were from one mice that could make a difference on the accuracy).

The prediction from the ResNets is based on the feature vectors which are placed at the end of convolutional layers. The model should produce similar feature vectors for the same input types and very different feature vectors for different input types. To verify the efficacy of our trained ResNets, the t-distributed stochastic neighbor embedding (t-SNE) method[48] was used, which has decreased the feature dimension from 256 to 3 for all the cell types. The features are plotted in the same coordinate space as shown in Fig. 4d, e. ResNet extracted features could help to distinguish B and T lymphocytes. Compared with principal component analysis (PCA) method, those features have made the boundaries between monocytes, granulocytes, and lymphocytes clearer (refer to Fig. S3 in **Supplementary Material**). To explore the cause of classification errors, some mistakenly classified cells are listed out. Apart from the morphological similarities between different cell types, the error could be also caused by the mislabeling in the ground truth dataset (refer to Fig. S5 in the **Supplementary Material**).

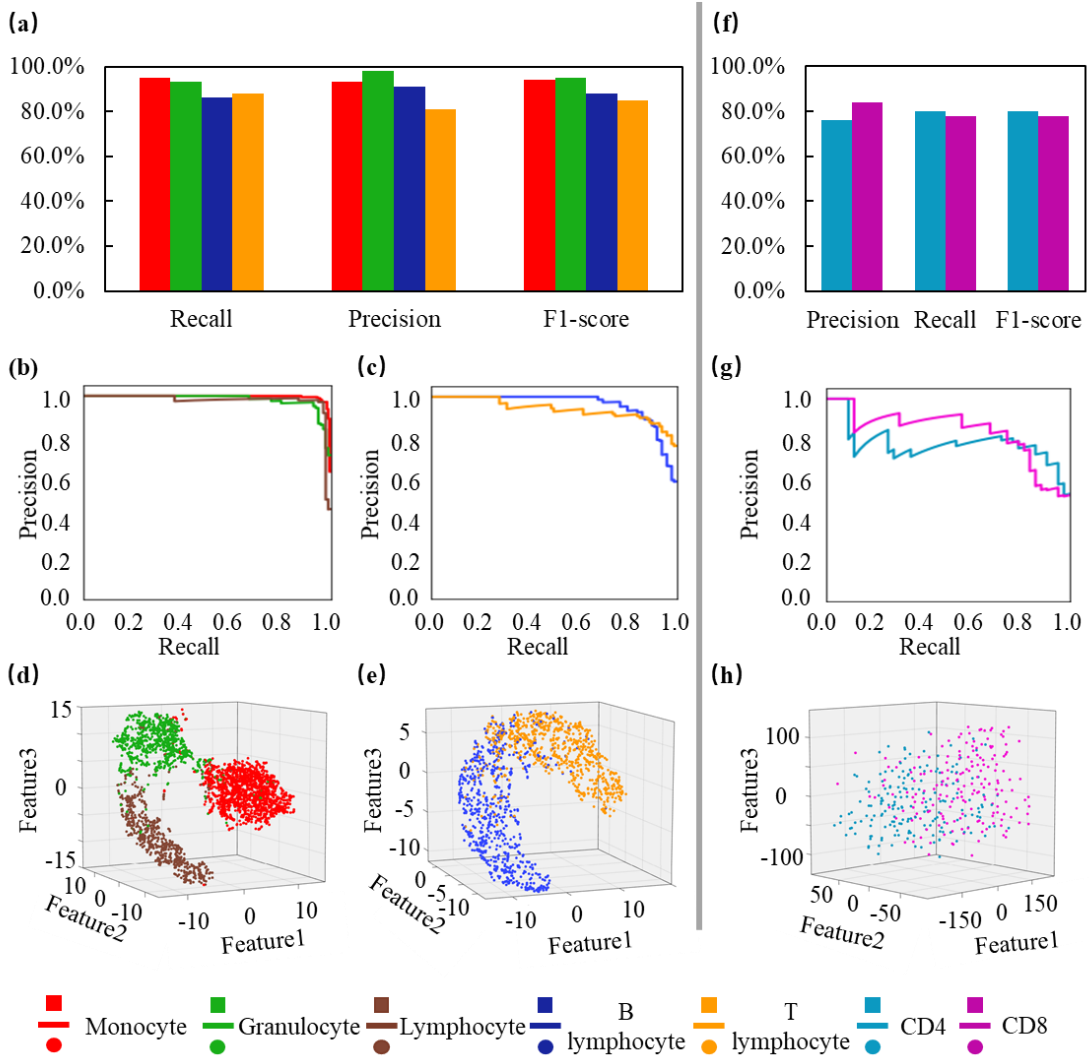

**Fig. 4 Classification results on monocytes (red), granulocytes (green), B lymphocytes (blue), T lymphocytes (orange), CD4 cells (light blue), and CD8 cells (purple).** **a**, Recall, precision, and F1-score for four types of leukocytes. **b**, Precision-recall curves for the monocyte-granulocyte-lymphocyte classifier. **c,** Precision-recall curves for the B-T lymphocyte classifier. **d,e,** T-SNE visualization of the feature extracted by the above two classifiers. **f,** Recall, precision, and F1-score for CD4 and CD8 cells. **g,** Precision-recall curve for the CD4-CD8 classifier. **h,** T-SNE visualization of the feature extracted by the CD4-CD8 classifier.

CD4 and CD8 cells are subtypes of T lymphocytes and have very similar morphological features[38] . Routine monitoring of CD4/CD8 cell ratio with point-of-care systems helps monitor immunodeficiency related diseases, e.g. acquired immunodeficiency syndrome (AIDS)[49,50]. Our proposed AI-powered platform has the potential to offer a unique approach in which the T cells can be virtually isolated and subtyped while also preserving them for subsequent immunophenotypic analysis. Moreover, such a platform can be expanded to visualize the immunological synapse due to its label-free attributes. We had previously demonstrated the use of QPM in identifying the activation state of CD8 cells in a contrast-free manner[23]. Building up on our previous study, we conjectured that our QPM can be used for differentiating CD4 and CD8 cells in a label-free manner. To test our hypothesis, we employed our AIRFIHA system on CD4 and CD8 cells from the same blood donor for both training and testing. The classification result is summarized in Fig. 4f-h. F1-scores of 80.4% and 77.5% for CD4 and CD8 cells are achieved, respectively (detailed values for recall, precision, and F1 scores are provided in Table S4). Compared with the F1-scores of 85.7% and 88.8% for CD4 and CD8 cells obtained by using 3D refractive maps[38], our preliminary results have a bit lower accuracy. The AUPRC values for CD4 and CD8 cells are 0.78 and 0.84, respectively. Using the t-SNE method, features are extracted from the CD4-CD8 classifier and plotted (Fig. 4h) for visualizing the differentiation capability. Our preliminary results show that our method has a basic differentiation capability for these two subtypes of T lymphocytes. The accuracy can be increased by using high volume of data and further tuning of our neural network.

**Cross-donor Validation**

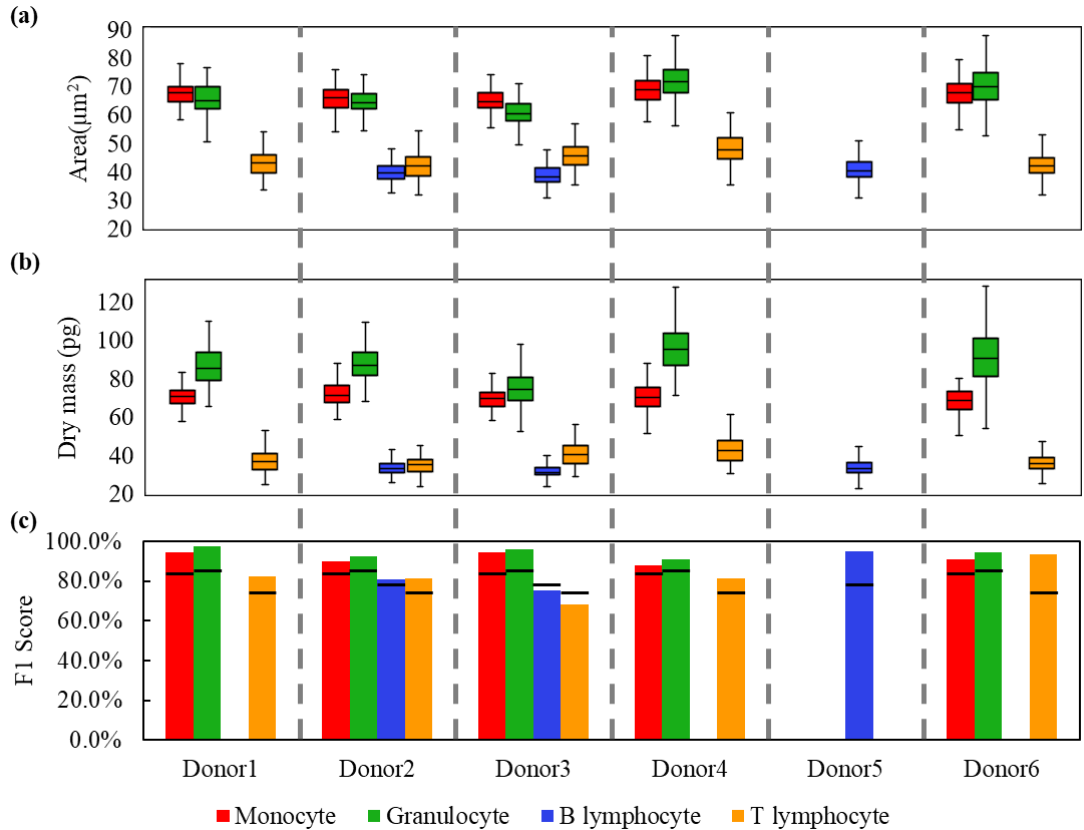

**Figure 5 a,** Comparison of leukocyte area distributions among different donors. Monocyte (red), Granulocyte (green), B lymphocyte (blue), T lymphocyte (orange) from 6 donors are included. **b,** Comparison of leukocyte dry mass distributions among different donors. **c,** F1-Score of leukocyte subtypes of each test donor from the cross-donor validation experiment. F1-scores obtained earlier using all six donors subtracting 10% are drawn with black lines for comparison.

As for real clinical applications, the blood test samples normally come from new individuals whose blood samples will not be known by our model. There could be variances in the morphological features of leukocytes of each type between different donors, depending on their age, health status, etc. [51,52] To verify whether such variances exist among our donors, we plotted the area and dry mass distributions for each donor (Fig. 5a,b), from which it was found out that there were indeed distribution differences between donors for certain leukocyte types (note that for several donors the distributions for certain types of leukocytes were missing). Since we had already acquired QPM images as a part of another work involving B cell leukemia, we decided not to measure the B cells from all the donors. This partly helped in ensuring that the extraction and subsequent QPM measurement from all the other leukocytes were completed within 3-4 hours of receiving the samples. In any case, we have more than 3 different donors for every leukocytes sample and we believe that it is sufficient for this proof-of-the principle investigation. The effect of such differences on the generalization of our model to the new donor was explored. For this purpose, the leukocyte samples from five donors were used for training and the leukocyte samples from the remaining donor was used for testing. This experiment was repeated by rotating the testing donor in the same group. The classification result is plotted in Fig. 5c (detailed numerical values are provided in Table S5). The result obtained earlier using all six donors (refer to Fig. 4a) is set as the reference for comparison. As for monocyte and granulocyte, all variances of the 10 tests are less than 0.1. For B lymphocytes only 1 out of 3 tests has variances larger than 0.1, while for T lymphocytes only 1 out of 5 tests has a variance larger than 0.1. The over-all average F1-scores for monocytes, granulocytes, and B and T lymphocytes are 91.6%, 94.3%, 83.7%, and 81.4%, respectively. Notable, the average F1-scores for the B-T lymphocyte classifier of the cascaded-ResNet are 84.1% and 85.9%. This have demonstrated a better accuracy when compared with an earlier work on the classification of B and T lymphocytes, where average F1-scores of 79.4% and 75.7% were obtained, respectively, using bright-field and dark-field imaging cytometry systems[36]. The cross-donor validation results have shown that our method has a high potential for clinical applications.

## Discussion

In this proof-of-concept study, the capability of AIRFIHA for label-free classification of leukocyte subpopulations has been demonstrated on human blood donors. With a well-designed neural network model, high information-content quantitative phase images, and a considerable amount of data collected from human blood donors, our AIRFIHA method has outperformed current reagent-free methods for the classification of granulocytes, monocytes, and B and T lymphocytes. Our preliminary result also shows that the detection accuracy of our method is not severely affected by different donors, thus indicating a potential for use in clinical settings. We have further demonstrated that AIRFIHA can differentiate CD4 and CD8 cells that are normally difficult to distinguish with label-free methods.

**Error Analysis.** It is important to note that our classification results rely on the accuracy of the separation kits used in this study to select the individual sets of leukocytes. We employed flow cytometry (refer to the details in "Methods") to measure the percentage population of the specific leukocytes after isolating them using the corresponding kits and the representative results from a donor are presented in **Supplementary** Fig. S4. These negative isolation kits have inherent inaccuracy that can adversely affect the classification results. However, compared with positive selection kits, negative selection kits could better maintain the original cell morphology for our label-free imaging modality, where the morphological attributes form the basis for classification.

**Result Evaluation.** We compared our result with other reported results using different detection/imaging principles, labeling methods, and experiment instruments, as shown in Table S8 in the **Supplementary Material**. AIRFIHA has a significantly improved accuracy when compared with the methods based on negative isolated leukocyte classification[53]. To a certain extent, our method benefits from the subtle differences in the refractive index maps of intracellular structure as encoded in the quantitative phase maps. For the classification of monocytes, granulocytes, and lymphocytes, our detection accuracy is slightly lower than the methods using positive fluorescence sorting or complicated purification methods [36,54,55]. It is possible that the negative selection kits have intrinsic lower accuracies in isolating leukocytes when compared with using positive kits, therefore reducing our classification accuracy. If there is a way to sort the leukocytes with higher accuracies without affect the original morphology states of cells, we expect to further increase the classification accuracy. For the classification of B and T lymphocytes, our result is better than bright and dark field microscopy based methods for the cross-donor validation experiments[36]. Our classification accuracy is also comparable with 3D QPM based methods that explore expensive and complex instrumentations (note that no human blood test and cross-donor validation have been carried in such methods so far) [38]. Notably, both mentioned methods are based on using positive leukocytes extraction methods. As for the classification of CD4 and CD8 cells, our classification accuracy is also compared with that obtained using 3D QPM methods[38].

**Further Improvement.** With the capability to differentiate very complex leukocyte types, AIRFIHA can provide more comprehensive information for potential disease diagnoses with simplified testing procedures. There are still ways to improve the detection accuracy of our system, such as improving the phase imaging resolution through synthetic aperture phase imaging method [56], deconvolution[57], and using 3D-resolved phase maps, preferably captured through a single image acquisition to avoid taking a large amount of data (such method has been recently made possible; a manuscript is under preparation by the authors). The other way to improve accuracy is to expand the dataset and upgrade the neural network model. With these improvements, we expect the generalization capability of our method can also be increased.

**Potential Applications.** Overall, our results show the potential of AIRFIHA as a fully automated, reagent-free, and high-throughput modality for differential diagnosis of leukocytes at point-of-care and in a clinical laboratory. Additional salient features of this platform include its single-shot measurement, small spatial footprint, and low cost. Of note, owing to its facile and simpler set-up, this platform can be combined with other modalities for blood cell investigation. For example, by combining it with microfluidic devices, AIRFIHA can conduct blood testing and analysis in a fully automated way. Importantly, the need for isolation kits is obviated and the leucocytes separated from blood using a routine centrifugation process can be directly subjected to the AIRFIHA to provide percentage population of leukocyte subtypes. One other example could be its integration with Raman spectroscopy that has been proposed for B lymphocytes acute lymphoblastic leukemia identification and classification[4]. While Raman spectroscopy provides biomolecular specificity, spontaneous Raman measurements are not feasible for clinical workflow requiring rapid diagnosis. Importantly, given the potential of the AIRFIHA platform in screening the B cells from other leucocytes, this QPM-based strategy can be used to screen the B lymphocytes where Raman measurements can be performed for B lymphocytes leukemia diagnosis. The combined QPM-Raman system obviates the need of any additional separation method to select B lymphocytes either from the blood or from the leucocyte mixtures for leukemia diagnosis in a label-free manner. Moreover, as

AIRFIHA involves a low-cost system that requires minimal sample preparation or chemical consumables, our AIRFIHA has a great potential to be used in point-of-care applications, resource-limited settings, or pandemic situations, e.g., COVID-19 pandemic, in view of a portable and low-cost QPM system recently demonstrated by us [58].

## Methods

**Fresh blood sample procurement.** The fresh blood samples from six anonymous healthy adult donors were purchased from StemCell Technologies (Vancouver, Canada) and all the experiments were conducted within 24 hours of blood extraction. The purchased blood samples contained ethylenediaminetetraacetic acid (EDTA) as the anti-coagulant.

**Leukocyte isolation from fresh blood.** Four types of leukocytes, namely monocytes, granulocytes, and B and T lymphocytes, were isolated from fresh blood samples using isolation kits from Stemcell Technologies. From each donor the amount of blood was in the 1-3 ml range, depending on the minimum volume requirement as per manufacturer's instruction for each leukocyte subpopulation. To isolate these four subpopulations, we used EasySep Direct Human Monocyte Isolation Kit, EasySep Direct Human Pan-Granulocyte Isolation Kit, EasySep Direct Human T Cell Isolation Kit, and EasySep Direct Human B Cell Isolation Kit (Stemcell Technologies Inc). These separation kits used immunomagnetic negative selection for isolating each specific leukocyte type from the whole blood sample. Two additional negative separation kits, i.e., EasySep Direct Human CD4+ T Cell Isolation Kit and EasySep Direct Human CD8+ T Cell Isolation Kit, were used for the isolation of CD4 and CD8 cells, respectively. Phosphate-buffered saline free from Ca++ and Mg++ (Gibco, Thermo Fisher Scientific) was used as the recommended medium for the EasySep Isolation kits. The isolation was carried out following the manufacturer's instructions with multiple cycles of mixing and incubation with the provided RapidSpheres and cocktail from the isolation kits. The final incubation yielded the isolated leukocytes in a 14 ml polystyrene round-bottom tube (Thermo Fischer

Scientific), which were centrifuged at 400g for 5 minutes. The cell pellet was resuspended in PBS before the cells were imaged.

**Flow cytometry analysis.** Flow cytometry was performed on the isolated leukocytes after the EasySep procedure to confirm the purity of the isolation. The viability of the leukocytes was checked with Acridine Orange and Propidium Iodide (AO/PI) staining (Invitrogen, Thermo Fischer Scientific) using a cell counter. The isolated leukocytes were counted and 50,000 of them were resuspended in cold PBS (Gibco, Thermo Fisher Scientific) at a density of $10^7$/ml. 100 ml of this cell suspension was added to each well in a 96 well plate. 1 µl of the required fluorophore-conjugated antibody was added to each well and incubated in the refrigerator for 20 mins. Anti-CD-14- PerCP was used for monocytes, Anti-CD-66b-FITC was used for granulocytes, Anti-CD-19- APC was used for B lymphocytes, and Anti-CD3- PE was used for T lymphocytes. The leukocytes were washed thrice with cold PBS and resuspended in 100 µl of cold PBS. The leukocytes were used for the flow cytometry analysis (MACSQuant Analyzer) and the data was analyzed with FlowJo software.

**Leukocyte sample preparation for quantitative phase imaging.** After the isolation of the leukocytes we suspended them in PBS solution and diluted five-ten times. DNase solution (1mg/ml) (Stemcell Technologies Inc) was added to the isolated cells to decrease the clumping and adsorption of protein fragments. Typically, 10 µl of the isolated cell suspension was sandwiched between two quartz coverslips and a secure seal spacer. Then, the sample was placed onto the sample-stage of the home-built system for quantitative phase imaging. We repeated this sample preparation procedure for collecting all the required phase images of leukocytes from each donor.

**Training of the classification model.** Phase maps of the leukocytes were obtained by cropping the phase images retrieved from the measured interferograms. Each phase map, containing one leukocyte, was then resized to 300x300 pixels to be used as the input of the network. In the training process, a 5-fold cross-validation method was used to tune the hyper-parameters, including network depth, batch size, etc. During the training, to ensure all leukocyte types were trained under the same condition (i.e., each type has the same number of training samples), the datasets of unbalanced leukocyte types were augmented by rotation, position shifting, and flipping. For the monocyte-granulocyte-lymphocyte classifier, B and T lymphocytes were treated as one type, i.e., lymphocytes, and then all granulocytes, monocytes and lymphocytes were used to train and test the classifier. Categorical cross-entropy loss and Adam optimizer (learning rate=$1 \times 10^{-3}$, $\beta_1$=0.9, $\beta_2$=0.999, learning rate decay=0)[59] were applied to optimize the model. In the end, the model with the best average validation accuracy was chosen as the final monocyte-granulocyte-lymphocyte classifier. For the B-T lymphocyte classifier, the dense layer of the obtained monocyte-granulocyte-lymphocyte classifier was first replaced with a new dense layer that has two outputs. All the B and T lymphocytes were used to fine-tune the entire network. Categorical cross-entropy loss and SGD optimizer (learning rate=$1 \times 10^{-3}$, learning rate decay=$1 \times 10^{-6}$, momentum=0.9)[60] were used. The network model with the best validation result was chosen as the final B-T lymphocyte classifier. By connecting these two network models, the final cascaded network model was obtained, from which the testing was conducted. The CD4-CD8 classifier was fine-tuned from the B-T lymphocyte classifier and trained and tested within the same donor. These frameworks were implemented with Tensorflow backend Keras framework and Python in the Microsoft Windows 10 operating system. The training was performed on a computer workstation, configured with an Intel i9-7900X CPU, 128 GB of RAM, and a Nvidia Titan XP GPU.

## Data availability

The data that support the findings of this study are available from the corresponding authors upon reasonable request.

## Acknowledgements

The authors are grateful to Prof. Pramod Srivastava for allowing us to use the flow cytometry facility and Dr. Sukrut Karandikar for his help with flow cytometer measurements. The authors are also thankful to Rosalie Bordett and Tiffany Liang for their help with the cell sorting experiments.

## Author contributions

R.P., and R.Z., conceived the original idea and directed the whole research work. R.Z., and R.P., designed and built the quantitative phase microscope. S.S. performed leukocyte isolation and imaging experiments following guidance from R.P. X.S., designed, implemented, and optimized the classification models and analyzed the results following guidance from R.Z. D.J., and X.Z., provided guidance on the optimization of the classification models. X.S., and R.Z., wrote the manuscript with contribution from all the authors.

## Competing interests

A US patent application has been filed based on this work.

## Additional information

**Supplementary Material** is available for this paper.

**Correspondence** and requests for materials should be addressed to R. P or R. Z.

Supplementary Materials for

# Artificial Intelligence Enabled Reagent-free Imaging Hematology Analyzer

Xin Shu,[1,#] Sameera Sansare,[2,3,#] Di Jin,[4] Xiangxiang Zeng,[5] Kai-Yu Tong,[1] Rishikesh Pandey,[3,6,*] Renjie Zhou,[1,7,*]

[1]Department of Biomedical Engineering, The Chinese University of Hong Kong, Shatin, New Territories, Hong Kong SAR, China
[2]Department of Pharmaceutical Sciences, University of Connecticut, Storrs, Connecticut, 06269, USA
[3]Connecticut Children's Innovation Center, University of Connecticut School of Medicine, Farmington, Connecticut, 06032, USA
[4]Computer Science and Artificial Intelligence Laboratory, Massachusetts Institute of Technology, Cambridge, Massachusetts 02139, USA
[5]School of Information Science and Engineering, Hunan University, Changsha 410076, China
[6]Department of Biomedical Engineering, University of Connecticut, Storrs, Connecticut, 06269, USA
[7]Shun Hing Institute of Advanced Engineering, The Chinese University of Hong Kong, Shatin, New Territories, Hong Kong SAR, China#: Authors contributed equally, Corresponding Authors' e-mail addresses: rishikesh.pandey@uconn.edu, rjzhou@cuhk.edu.hk

## 1. Diffraction phase microscopy system

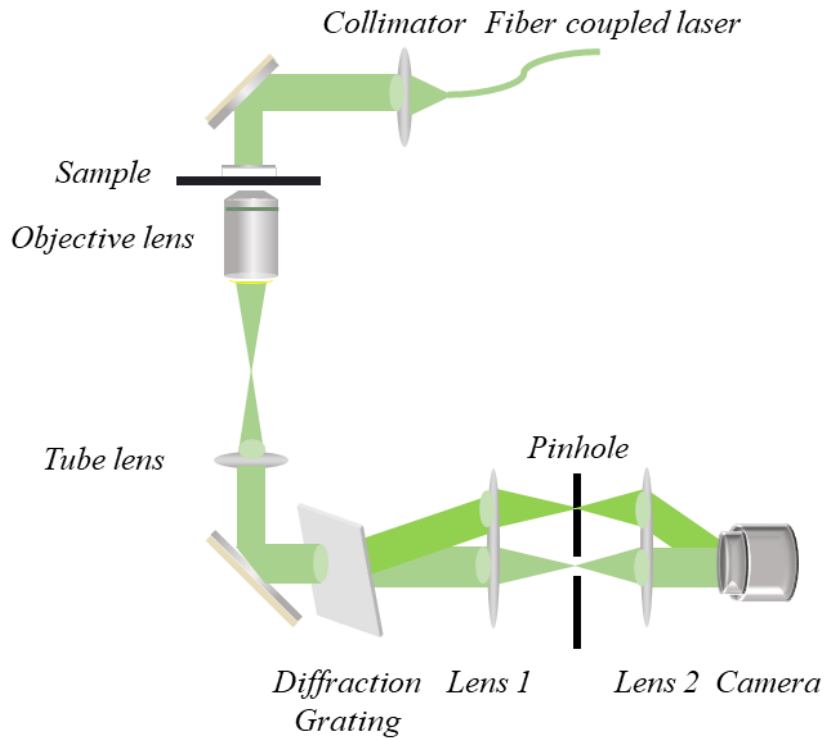

Fig. S1. The schematic design of the diffraction phase microscopy system.

Diffraction phase microscopy (DPM) is a common-path quantitative phase microscopy (QPM) method that allows for highly sensitive measurement of cell morphology with nanometer-scale sensitivity [1]. As only one interferogram is needed to obtain a wide-field phase map, high-speed image acquisition is possible with DPM. We have recently developed a portable DPM system with a low-cost to enable a broader adoption [2]. The DPM system, as illustrated in Fig. S1, is used to measure the phase maps of the leukocytes. A 532 nm laser (Gem 532, Laser Quantum) is used as the illumination source for the system. The collimated laser beam first passes through the sample, and then the sample scattered field is collected by a water dipping objective lens with numerical aperture (NA) of 1.1 (LUMFLN60XW, Olympus). After that, the sample beam goes through a tube lens and forms an intermediate image at its back focal plane. A diffraction grating, placed at the intermediate image plane, produces multiple copies of the sample image. Two of the diffraction orders are selected by a subsequential $4f$ system formed by lens 1 and lens 2. The $1^{st}$ order beam is filtered down to a DC beam (or reference beam) through a 10 μm diameter pinhole filter, placed at the Fourier plane of lens 1. The $0^{th}$ order beam passes the $4f$ system without any filtering as serves as the signal beam. At the final imaging plane after lens 2, these two beams interfere with each other and form an interferogram which is then captured by a USB camera (FL3-U3-13Y3M-C, Pointgrey). The imaging system has a total magnification of around 100, a lateral resolution of around 590 nm according to the Rayleigh criterion, and a field of view of 61 μm x 49 μm.

## 2. Quantitative phase image processing

The phase image processing mainly consists of phase retrieval [1] and segmentation, as shown in Fig. S2. A Fourier transform is first performed over the raw interferogram (first column in Fig. S2), and then a bandpass filter is used to select the +1 or -1 order signal. After that, the selected signal is shifted back to the origin of the frequency spectrum. An inverse Fourier transform is performed to obtain the complex sample field. Meanwhile, another interferogram taken in the sample-free region is used as the calibration image and the same processing is conducted to obtain the complex calibration field. Then the calibration complex field is divided from the sample complex field to obtain the calibrated sample field, from which the sample phase map is obtained. Subsequently, a phase unwrapping procedure is added to unwrap the sample phase map. Finally, we flatten and zero the phase map by removing the background tilt and subtracting the background phase value. Representative phase images for each major leukocyte type are shown in the second column in Fig. S2. After obtaining the phase images, we select each individual cells with a segmentation algorithm [3] and create cell phase maps (third column in Fig. S2). To ensure the same size for all the cell phase maps, we paste each cell phase map on a fixed-size template.

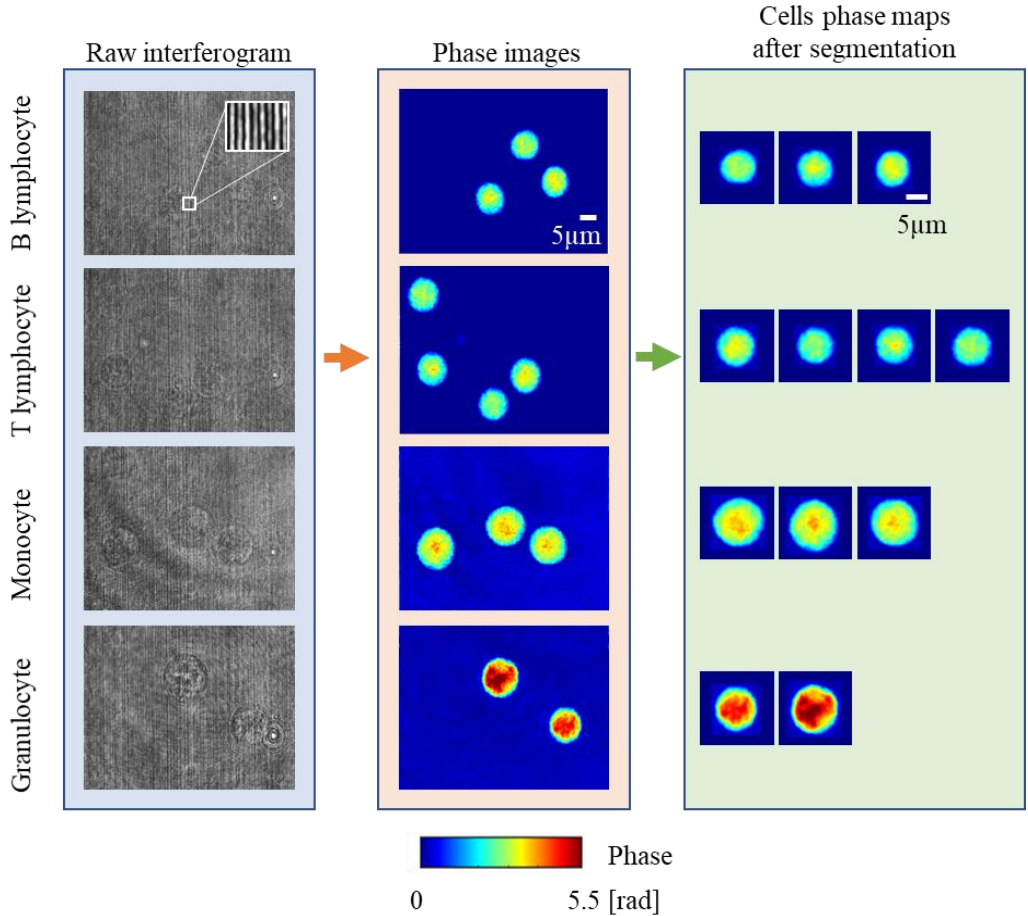

Fig. S2. Illustration of the quantitative phase image processing steps. The phase retrieval step is first performed over the raw interferograms (representative interferograms for each major leukocyte type are shown) to obtain the phase images. In the second step, a segmentation algorithm is used to select individual cells and create their phase maps.

## 3. Principal component analysis (PCA)

We first reshape each image with size of 300x300 into a 1x90000 sequence and then use the principal component analysis (PCA) [4] method to decrease the dimension from 90000 to 256. At last, by using the t-distributed stochastic neighbor embedding (t-SNE) method [5], we visualize the PCA extracted features in a 3-D plot.

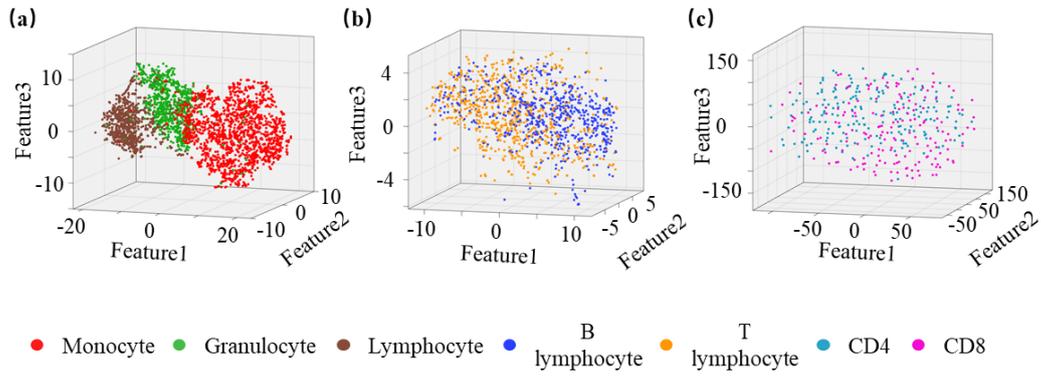

Fig. S3. Visualization of the features extracted by PCA using the T-SNE method. **a**, Visualization of PCA features of monocytes, granulocytes, and lymphocytes. **b**, Visualization of PCA features of B and T lymphocytes. **c**, Visualization of PCA features of CD4 and CD8 cells.

## 4. Flow cytometry measurements

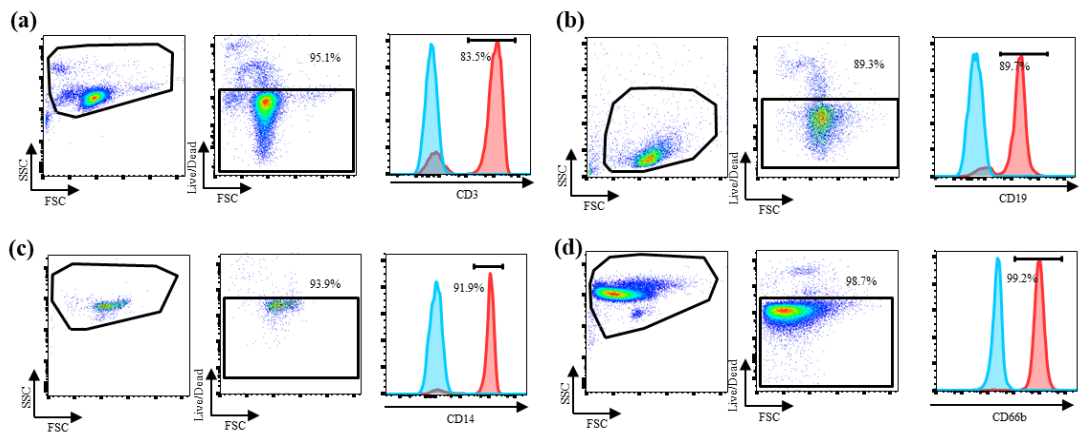

Fig. S4. Flow cytometry analysis for the purity of isolated leukocytes showing the polygonal gating for live leukocytes along with the fluorophore-conjugated antibodies for each leukocyte type. **a**, Anti-CD3-PE for T lymphocytes. **b**, Anti-CD-19-APC for B lymphocytes. **c**, Anti-CD-14-PerCP for monocytes. **d**, Anti-CD-66b-FITC for granulocytes. FSC: forward scatter; SSC: side scatter.

The flow cytometry results for assessing the purity of isolated leukocytes are illustrated in Fig. S4. The percentage population for T lymphocytes, B lymphocytes, monocytes, and granulocytes in representative isolated leukocyte samples were 83.5%, 89.7%, 91.9%, and 99.2%, respectively.

# 5. Detailed leukocyte classification results

| Table S1. Classification result from the monocyte-granulocyte-lymphocyte classifier | | | | | | |
|---|---|---|---|---|---|---|
| | **Predicted type** | | | | **Recall** | **F1-score** |
| **Label type** | | Lymphocyte | Monocyte | Granulocyte | | |
| | Lymphocyte | 197 | 2 | 1 | 98.5% | 97.7% |
| | Monocyte | 4 | 95 | 1 | 95.0% | 94.0% |
| | Granulocyte | 2 | 5 | 93 | 93.0% | 95.4% |
| **Precision** | | 97.0% | 93.1% | 97.9% | Accuracy | 96.3% |

| Table S2. Classification result from the B-T lymphocyte classifier | | | | | |
|---|---|---|---|---|---|
| | **Predicted type** | | | **Recall** | **F1-score** |
| **Label type** | | B lymphocyte | T lymphocyte | | |
| | B lymphocyte | 86 | 14 | 86.0% | 88.2% |
| | T lymphocyte | 9 | 91 | 91.0% | 88.8% |
| **Precision** | | 90.5% | 86.7% | Accuracy | 88.5% |

| Table S3. Summarized classification result from the cascaded-ResNet | | | | | | | |
|---|---|---|---|---|---|---|---|
| | **Predicted type** | | | | | **Recall** | **F1-score** |
| **Label type** | | Monocyte | Granulocyte | B lymphocyte | T lymphocyte | | |
| | Monocyte | 86 | 14 | 0 | 0 | 86.0% | 88.2% |
| | Granulocyte | 9 | 88 | 2 | 1 | 88.0% | 84.6% |
| | B lymphocyte | 0 | 4 | 95 | 1 | 95.0% | 94.0% |
| | T lymphocyte | 0 | 2 | 5 | 93 | 93.0% | 95.4% |
| **Precision** | | 90.5% | 81.5% | 93.1% | 97.9% | Accuracy | 90.5% |

| Table S4. Classification result from the CD4-CD8 classifier from one donor | | | | | |
|---|---|---|---|---|---|
| | Predicted type | | | Recall | F1-score |
| Label type | | CD4 | CD8 | | |
| | CD4 | 37 | 6 | 86.0% | 80.4% |
| | CD8 | 12 | 31 | 72.1% | 77.5% |
| Precision | | 75.5% | 83.8% | Accuracy | 79.1% |

| Table S5. Cross-donor testing result with cascaded-ResNet | | | | |
|---|---|---|---|---|
| Test donor | B lymphocyte (F1-score) | T lymphocyte (F1-score) | Monocyte (F1-score) | Granulocyte (F1-score) |
| 1 | N/A | 82.5% | 94.4% | 97.5% |
| 2 | 80.9% | 81.2% | 90.2% | 92.5% |
| 3 | 75.3% | 68.5% | 94.5% | 96.0% |
| 4 | N/A | 81.3% | 87.8% | 90.9% |
| 5 | 94.9% | N/A | N/A | N/A |
| 6 | N/A | 93.5% | 91.3% | 94.6% |

*N/A represents when such data is unavailable, or the dataset is too small to have a statistical significance.

| Table S6. Cross-donor testing result from the monocyte-granulocyte-lymphocyte classifier | | | |
|---|---|---|---|
| Test donor | Lymphocyte (F1-score) | Monocyte (F1-score) | Granulocyte (F1-score) |
| 1 | 96.8% | 94.4% | 97.5% |
| 2 | 98.2% | 90.2% | 92.5% |
| 3 | 97.4% | 94.5% | 96.0% |
| 4 | 90.0% | 87.8% | 90.9% |
| 5 | 99.9% | N/A | N/A |
| 6 | 96.0% | 91.3% | 94.6% |

*N/A represents when such data is unavailable, or the dataset is too small to have a statistical significance.

| Table S7. Cross-donor testing result from the B-T lymphocyte classifier | | |
|---|---|---|
| **Test donor** | **B lymphocyte (F1-score)** | **T lymphocyte (F1-score)** |
| 1 | N/A | 85.5% |
| 2 | 81.7% | 82.9% |
| 3 | 75.9% | 71.5% |
| 4 | N/A | 92.0% |
| 5 | 94.9% | N/A |
| 6 | N/A | 97.6% |

*N/A represents when such data is unavailable, or the dataset is too small to have a statistical significance.

## 6. Examples of misclassified leukocytes

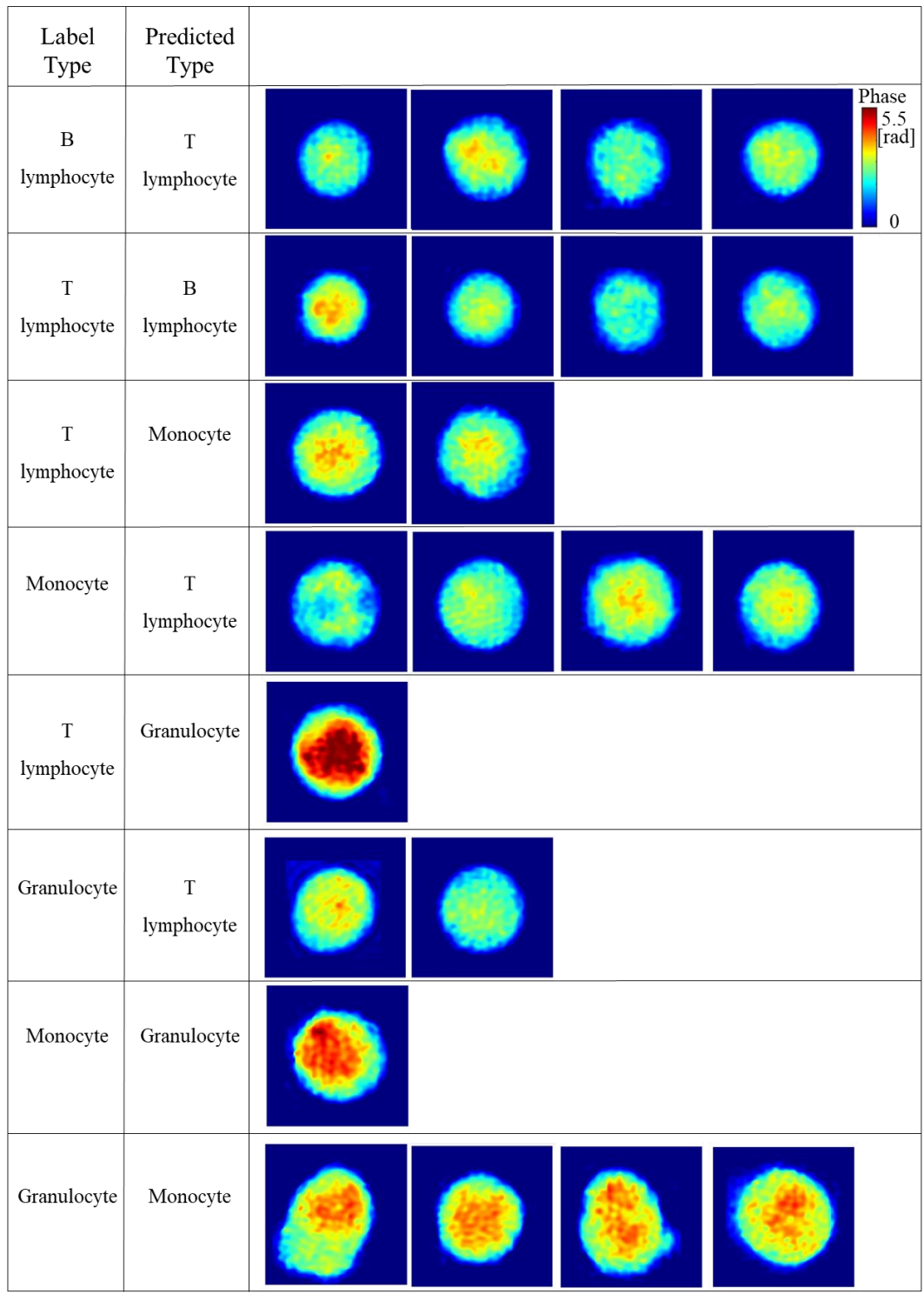

Fig. S5. Selected examples of misclassified leukocytes.

To explore the cause of the misclassification, we show the phase maps of several selected misclassified leukocytes and their corresponding actual types and predicted types (Fig. S5). We suspect some of the misclassifications might be induced by mislabeling. For example, the leukocyte labelled as a T lymphocyte but predicted as a granulocyte has large phase values and a large area, which are not in accordance with typical features of T lymphocytes.

## 7. Comparison with results from existing methods

| Table S8. Comparison of AIRFIHA with existing methods on classification accuracy | | | | | | |
|---|---|---|---|---|---|---|
| **Method** | **Labeling method** | **Sample type** | **Cross-validation** | **Result** | | |
| | | | | Monocyte | Granulocyte | Lymphocyte |
| Bright and dark field microscope [6] | Fluorescence cytometry | Human | Yes | 96.0% | 96.4% (Neutrophil) 96.3% (Eosinophil) | 96.9% |
| Lens-free holography [7] | Fluorescence cytometry | Human | No | 98.4% | 98.5% | 98.7% |
| Lens-free holography [8] | Negative Immunomagnetic depletion | Human | No | 91.0% | 92.8% | 85.5% |
| Third harmonic generation microscope [9] | Density centrifugation + Scattered light cytometry + Negative fluorescence cytometry | Human | No | 97.5% | 97.5% | 98.0% |
| **AIRFIHA** | Negative Immunomagnetic depletion | Human | No | 94.0% | 95.4% | 97.7% |
| AIRFIHA | Negative Immunomagnetic depletion | Human | Yes | 91.6% | 94.3% | 96.4% |
| | | | | | B lymphocyte | T lymphocyte |
| Bright and dark field microscope [6] | Fluorescence cytometry | Human | Yes | | 79.4% | 75.7% |
| Optical diffraction tomography [10] | Fluorescence cytometry | Mice | No | | 88.4% | 90.9% |
| **AIRFIHA** | Negative Immunomagnetic depletion | Human | No | | 88.2% | 88.8% |
| **AIRFIHA** | **Negative Immunomagnetic depletion** | **Human** | **Yes** | | **84.1%** | **85.9%** |
| | | | | | CD4 | CD8 |
| Optical diffraction tomography [10] | Fluorescence cytometry | Mice | No | | 85.7% | 88.8% |
| **AIRFIHA** | Negative Immunomagnetic depletion | Human | No | | 80.4% | 77.5% |